%% file: main.tex
\documentclass[sigconf]{acmart}
\microtypesetup{protrusion=false, expansion=false, activate={true,nocompatibility}, patch=none}
\usepackage{xcolor}
\usepackage{balance}
\usepackage{float}
\usepackage{bbding}
\usepackage{hyperref}
\usepackage{fontawesome}
\usepackage{setspace} 
\usepackage{todonotes}
\usepackage{pifont}
\usepackage{graphicx}

\DeclareRobustCommand{\circlednum}[1]{
  \begingroup
    \setbox0=\hbox{X}
    \dimen0=\ht0
    \raisebox{-0.15ex}{\resizebox{!}{1.2\dimen0}{\ding{\numexpr181 + #1\relax}}}
  \endgroup
}
\newcommand{\rsastage}[1]{\circlednum{#1}~(Fig.~\ref{fig:overview})}

\input{commands}

 \newcommand{\citenanopore}{menestrina_ionic_1986,cherf_automated_2012,laszlo_decoding_2014,deamer_three_2016,kasianowicz_characterization_1996,meller_rapid_2000,stoddart_single-nucleotide_2009,derrington_nanopore_2010,song_structure_1996,walker_pore-forming_1994,wescoe_nanopores_2014,lieberman_processive_2010,bezrukov_dynamics_1996,stoddart_nucleobase_2010,ashkenasy_recognizing_2005,stoddart_multiple_2010,bezrukov_current_1993,zhang_single-molecule_2024}
\newcommand{\citebasecallnanodnn}{cavlak_targetcall_2024,xu_fast-bonito_2021,peresini_nanopore_2021,boza_deepnano_2017,boza_deepnano-blitz_2020,oxford_nanopore_technologies_dorado_2024,oxford_nanopore_technologies_guppy_2017,lv_end--end_2020,singh_rubicon_2024,zhang_nanopore_2020,xu_lokatt_2023,zeng_causalcall_2020,teng_chiron_2018,konishi_halcyon_2021,yeh_msrcall_2022,noordijk_baseless_2023,huang_sacall_2022,miculinic_mincall_2019}

\newcommand{\citesignalanalysis}{bao_squigglenet_2021,loose_real-time_2016,zhang_real-time_2021,kovaka_targeted_2021,senanayake_deepselectnet_2023,lindegger_rawalign_2023,firtina_rawhash_2023,firtina_rawhash2_2024,shih_efficient_2023,sadasivan_rapid_2023,dunn_squigglefilter_2021,Shivakumar2024,sadasivan_accelerated_2024,gamaarachchi_gpu_2020,samarasinghe_energy_2021,firtinac_2024_rawsamble}

\copyrightyear{2025}
\acmYear{2025}
\setcopyright{cc}
\setcctype{by}
\acmConference[BCB '25]{Proceedings of the 16th ACM International Conference on Bioinformatics, Computational Biology, and Health Informatics}{October 11--15, 2025}{Philadelphia, PA, USA}
\acmBooktitle{Proceedings of the 16th ACM International Conference on Bioinformatics, Computational Biology, and Health Informatics (BCB '25), October 11--15, 2025, Philadelphia, PA, USA}
\acmDOI{10.1145/3765612.3767302}
\acmISBN{979-8-4007-2200-4/2025/10}
\begin{document}

\title{RawBench: A Comprehensive Benchmarking Framework \\
for Raw Nanopore Signal Analysis \omnd{Techniques}}
\thispagestyle{firstpagestyle}
\author{Furkan Eris}
\affiliation{
  \institution{ETH Zurich}
  \city{Zurich}
  \country{Switzerland}
}
\email{furkan6olm@gmail.com}

\author{Ulysse McConnell}
\affiliation{
  \institution{ETH Zurich}
  \city{Zurich}
  \country{Switzerland}
}
\email{umcconnell@student.ethz.ch}

\author{Can Firtina}
\authornote{Corresponding author\omst{s}}
\affiliation{
  \institution{University of Maryland, College Park}
  \state{MD}
  \country{USA}
}
\email{firtina@umd.edu}

\thispagestyle{firstpagestyle}
\author{Onur Mutlu}
\authornotemark[1]
\affiliation{
  \institution{ETH Zurich}
  \city{Zurich}
  \country{Switzerland}
}
\email{omutlu@gmail.com}

\begin{abstract}
Nanopore sequencing technologies continue to advance rapidly, offering critical benefits such as real-time analysis, the ability to sequence extremely long DNA fragments \omst{(}up to millions of bases in a single read\omst{), }and the option to selectively stop sequencing a molecule before completion. Traditionally, the raw electrical signals generated during sequencing are converted into DNA sequences through a process called \emph{basecalling}, which typically relies on large neural network models. While accurate, these models are computationally intensive and often require high-end GPUs to process the vast volume of raw signal data. This presents a significant challenge for real-time processing, particularly on edge devices with limited computational resources, ultimately restricting the scalability and deployment of nanopore sequencing in resource-constrained settings. \omnd{R}aw signal analysis has emerged as a promising alternative to these resource-intensive approaches. While attempts have been made to benchmark conventional basecalling methods, existing evaluation frameworks \reve{1) overlook raw signal analysis techniques, 2) lack the flexibility to accommodate new \omst{raw signal analysis} tools \omst{easily}, and 3) fail to include the latest improvements in nanopore datasets}. Our goal is to provide an \emph{extensible} benchmarking \omst{framework} that enables designing and comparing new methods for raw signal analysis. To this end, we introduce RawBench, the \emph{first} flexible framework for evaluating raw nanopore signal analysis \omst{techniques}. RawBench provides modular evaluation of three core pipeline components: 1) reference genome encoding (using different pore models), 2) signal encoding (through various segmentation methods), and 3) representation matching (via different data structures). We \emph{extensively} evaluate raw signal analysis techniques in terms of 1) quality and performance for read mapping, 2) \reve{quality} and performance for read classification, and 3) \reve{quality} of raw signal analysis-assisted basecalling. 
Our evaluations show that raw signal analysis can achieve competitive \reve{quality} while significantly reducing resource requirements, particularly in settings where real-time processing or edge deployment is necessary. 
\end{abstract}

\maketitle

\section{Introduction}
\label{intro}
\omst{N}anopore sequencing\omrd{~\cite{\citenanopore}} has revolutionized genomics by enabling the analysis of exceptionally long DNA molecules \rev{up to 4 million bases~\cite{jain_nanopore_2018,pugh_current_2023,senol_cali_nanopore_2019,jain_klobases_whales_2021}. To sequence each base}, nanopore devices measure ionic current changes as \rev{raw electrical signals while} nucleic acids pass through \omst{nanoscale} biological pores,  \rev{called \emph{nanopores}}~\cite{deamer_three_2016}. This approach offers two \omst{major} benefits: 1) real-time sequencing decisions \revb{without having to fully sequence every read, a technique known as \emph{adaptive sampling}, \reve{to reduce} sequencing time and cost ~\cite{sadasivan_rapid_2023,senanayake_deepselectnet_2023,dunn_squigglefilter_2021,zhang_real-time_2021,kovaka_targeted_2021,bao_squigglenet_2021,Ulrich2022,Payne2020}}, and \rev{2)} \reve{ natively providing richer information, such as epigenetic modifications (e.g., methylation patterns)}\rev{~\cite{flynn_evaluation_2022, Ahsan2024, Doshi2025,Liu2021}}.

\rev{There are two main approaches to analyzing nanopore sequencing data. The most widely adopted method is basecalling where \omnd{raw} electrical signals are translated into nucleotide sequences \omnd{(i.e., A, C, G, T)}\reve{, called reads,} using computationally intensive deep learning models~\cite{\citebasecallnanodnn}. This approach has the advantage of producing accurate reads that are compatible with existing bioinformatics pipelines. However, basecalling requires processing large segments of signal data~\cite{zhang_real-time_2021}, \revb{which increases latency and memory usage,} and typically demands high-performance computing resources\omst{~\cite{shih_efficient_2023, senol_cali_nanopore_2019, singh_rubicon_2024}}. These requirements \omst{make real-time analysis challenging, i.e., processing and interpreting sequencing data as it is generated \cite{zhang_real-time_2021}. They also hinder portable sequencing, i.e., the use of field-deployable DNA/RNA sequencing devices such as Oxford Nanopore's MinION, in resource-constrained environments \cite{shih_efficient_2023}.} \revb{While basecalling pipelines can be adapted for real-time use, doing so often requires \omst{reduction in quality} and reliance on specialized or high-end hardware to meet latency constraints\omst{~\cite{ont_x,mobile_dna,fpga3rd_dna,helix,brawl,cimba,genpip, mars}.}}} 

\rev{An emerging alternative \reve{that addresses these limitations is to analyze raw signals \emph{directly} without basecalling} \omrd{~\cite{\citesignalanalysis}}. \reve{By bypassing the translation step \omst{of electrical signals to text}, direct analysis can significantly reduce computational overhead, lower memory usage, and enable faster decision-making. This is particularly advantageous in scenarios like} adaptive sampling, where sequencing decisions must be made in real time \cite{zhang_real-time_2021}.} 

\reve{Raw signal analysis\omst{ offers} unique opportunities to extract richer biological insights directly from raw data. Over the past years, a growing number of tools have been developed in this direction each advancing the state-of-the-art in terms of \reve{quality}, speed, and resource usage \omrd{\cite{\citesignalanalysis}}. To enable fair comparison\omst{s} and better assessment of trade-offs, several benchmarking frameworks \omst{for raw \omnd{nanopore} signal analysis} have been proposed \crfinal{\cite{cheng_nanobaselib_2024,pages-gallego_comprehensive_2023}}.}

\reve{However, existing benchmarking solutions lack three critical capabilities: (1) the ability to support \omnd{both basecalled read and raw signal analysis (RSA)}, (2) the flexibility to incorporate newly developed methods \omst{targeting individual steps of raw signal analysis}, and (3) access to standardized datasets that reflect the latest improvements in nanopore \omst{sequencing} technolog\omst{y}. \omst{These shortcomings hinder} (1) \omst{\emph{comprehensive}} evaluation of raw signal analysis techniques, (2) high-resolution assessment \omst{that allows mixing and matching techniques from different tools}, and (3) fair comparison of tools across different nanopore chemistries and raw signal behavior.}

\reve{We identify three key challenges that underlie these \omst{shortcomings}. First, prior benchmarking efforts\crfinal{~\cite{cheng_nanobaselib_2024,pages-gallego_comprehensive_2023}} adopt a task-focused approach, often focusing on a specific analysis goal, e.g., basecalling or modification detection, based on the capabilities of a small number of tools. This limits broader insight, as the chosen task may not reflect the generalizable strengths of different \omst{raw signal analysis} techniques. Second, \omst{prior} benchmarking solutions are designed as monolithic systems to ease implementation\omst{, and hence} the modularity needed to flexibly incorporate and evaluate new methods. Third, there is limited availability of comprehensive and standardized raw signal datasets, due to both the default discarding of the raw nanopore signal data during basecalling (e.g., by ONT’s MinKNOW software) and the general lack of public sharing~\cite{Mencius2025}.}

\omst{\textbf{Our goal}} is to enable comprehensive and extensible benchmarking \omst{of} raw nanopore signal analysis methods that support both basecall\crfinal{ed} and \omnd{RSA} approaches on large, representative datasets. To this end, we propose RawBench, the \emph{first} modular raw nanopore signal analysis benchmarking \omst{framework} designed to address key limitations of existing \omst{benchmarking} frameworks.

\reve{We demonstrate the capabilities of RawBench through an extensive benchmarking study\omst{,} evaluating \omst{two} commonly used tasks\omst{, i.e., read mapping and read classification,} that \crfinal{can} leverage raw signal analysis. Our study systematically evaluates \omst{thirty} diverse algorithmic strategies across a \omst{wide} range of \omst{four} datasets and \omst{two} experimental conditions. Our key contributions include:}

\begin{itemize}
    \item \reve{We benchmark \crfinal{3}0 unique method combinations for raw signal analysis, exploring key design trade\omst{-}offs and scaling \omst{characteristics} across tasks and sequencing conditions. \omst{Our comprehensive structure} provides insights critical for real-time use and resource-constrained environments.}
    
    \item \reve{We enable flexible integration of new \omst{raw signal analysis} methods, facilitating fine-grained comparisons and high-resolution method development, addressing the inflexibility of prior benchmarking tools and establishing the infrastructure necessary for \omst{well-documented} \omnd{and open} progress in \crfinal{raw} signal analysis.}
    
    \item \reve{We incorporate curated, up-to-date raw nanopore datasets spanning multiple species and sequencing chemistries, providing a robust foundation for fair and reproducible evaluations.}
    
    \item \reve{We \omst{fully open source} RawBench’s codebase, datasets, and benchmark results to foster transparency and accelerate future research in raw nanopore signal analysis.} \revf{\href{https://github.com/CMU-SAFARI/RawBench}{\faGithub}}
\end{itemize}

\rev{\section{Background\omnd{, Related Work, Our Goal}}}
\label{background}
\rev{Nanopore sequencing uniquely enables \omnd{\emph{adaptive sampling}}, a technique that \revb{decides in real time whether to continue or stop reading a DNA molecule based on its initial signal \omst{reading}. This lets the system focus on molecules that are likely to be important (such as genes linked to a disease) while skipping irrelevant ones. Unlike traditional \omst{enrichment} methods that require special lab preparation, adaptive sampling performs this selection during sequencing\reve{,} saving} time and cost ~\cite{Edwards2019, Ulrich2022}. \revb{When combined with the portability of nanopore sequencers}, \omst{adaptive sampling} \omst{enables} field applications such as rapid pathogen surveillance~\cite{bloemen_development_2023, BronzatoBadial2018}. However, real-time decision-making \omst{aspect of adaptive sampling} imposes stringent latency constraints, as decisions must occur \revb{fast enough} to \omst{fully utilize} pore throughput~\cite{zhang_real-time_2021,firtina_rawhash_2023}, demanding lightweight methods on} resource-limited edge devices \cite{shih_efficient_2023}.

\rev{Basecalling \omnd{is} the dominant approach \omnd{to analyzing} nanopore signal\omnd{s}, where deep neural networks (e.g., convolutional / recurrent architectures) translate raw ionic current signals into \omnd{basecalled} nucleotide sequences \omnd{(i.e., A, C, G, T)}\omnd{~\cite{\citebasecallnanodnn}}. State-of-the-art \omst{basecallers} like Dorado~\cite{oxford_nanopore_technologies_dorado_2024} achieve high \reve{quality} but rely on complex\omst{, resource-intensive architectures that create bottlenecks} for downstream genomic analyses. This overhead arises from processing long signal traces\reve{,} often millions of data points per read\reve{,}  using \revb{neural networks with substantial redundancy, as evidenced by \omst{a} recent work showing that up to 85\% of model weights in such models can be pruned without \omnd{significant} \reve{quality} loss} \cite{singh_rubicon_2024}. Consequently, basecalling bottlenecks real-time portable sequencing \omst{\cite{dunn_squigglefilter_2021, cimba, shih_efficient_2023}}, motivating efforts to accelerate performance via hardware and algorithmic optimizations\omnd{~\cite{Mutlu2023,Alser2022,Alser2020}}.}

\rev{To accelerate basecalling, \omst{several works} explore FPGAs \omrd{\cite{ont_x,mobile_dna,fpga3rd_dna,fpgabased_dna_accel,fpga_dna_riscv, singh_rubicon_2024}} and PIM-based solutions \cite{helix,brawl, genpip, cimba}, leveraging parallelism \omst{and eliminating data movement overhead} to reduce latency and energy \omst{consumption}. GenPIP \cite{genpip} \omnd{concurrently} executes basecalling and read mapping\reve{, the process of aligning sequenced reads to a reference genome,} in\omnd{side} memory to minimize data movement and redundant computation. CiMBA \cite{cimba} advances the field with a compact compute-in-memory accelerator and analog-aware networks for on-device basecalling. While effective, these solutions require specialized hardware\reve{, limiting their adaptability to changes in the sequencing technology and deployment \omst{scenario}}.}

\rev{A promising alternative to basecalling is \omnd{RSA}, which operates directly on electrical signal\omnd{s} without translating \omnd{them} into \omnd{reads}\omnd{~\cite{zhang_real-time_2021, firtina_rawhash_2023, firtina_rawhash2_2024}}. These techniques typically involve three \reve{distinct} stages\reve{.} (1) \reve{In the} \omnd{\emph{reference encoding}} \reve{stage}, a genome sequence is converted into expected \omnd{corresponding} \revb{electrical} signal\omnd{s} using a pore model \revb{that predicts current levels for each DNA k-mer\omst{, i.e., short nucleotide sequences of length $k$}.} (2) \reve{In the} \omnd{\emph{signal \omnd{segmentation}}} \reve{stage}, the continuous nanopore current \reve{is divided} into discrete segments corresponding to individual k-mers. This segmentation step explicitly extracts meaningful \omst{segments} from noisy raw \omst{signal} data using statistical or learn\omnd{ing} methods, unlike basecalling where segmentation is \omst{performed} implicit\omst{ly} in neural networks\reve{.} (3) \reve{In the} \omnd{\emph{representation matching}} \reve{stage}, \revb{\reve{the} segmented signals are compared against the encoded reference using different matching algorithms to find similarities despite \omst{the} noise \omst{in the signal data}. More details on methods addressing each of these stages can be found in Section \ref{framework}}.}

\rev{\revb{Beyond algorithmic \omnd{developments for RSA},} recent works demonstrate the \omnd{benefits} of real-time raw signal processing on specialized hardware. SquiggleFilter \cite{dunn_squigglefilter_2021} uses an ASIC to filter irrelevant reads \omnd{before} basecalling, eliminating unnecessary and costly basecalling and enabling fast \omnd{pathogen} detection. HARU \cite{shih_efficient_2023} introduces an FPGA-based accelerator for real-time adaptive sampling in resource-constrained environments. MARS ~\cite{mars} adopts a storage-centric Processing-In-Memory (PIM) \omnd{approach}\omst{, i.e., where computation is performed directly inside or near memory,} to accelerate raw signal genome analysis, combining filtering, quantization, and in-storage execution to achieve up to 28× speedup and 180× energy savings over software baselines. \reve{Together, these efforts underscore the growing interest in hardware-software co-design for raw signal analysis and unlock capabilities for the diverse and dynamic landscape of nanopore sequencing applications}.}

\rev{As RSA grows\reve{,} particularly in latency- and energy-sensitive settings\reve{,} there is an increasing need for fair, systematic evaluation of emerging tools and techniques. Benchmarking becomes critical not only to measure performance but also to understand the trade-offs between \omst{different design} choices. However, existing benchmarking efforts have limitations. While NanoBaseLib~\cite{cheng_nanobaselib_2024} and basecalling benchmarks~\cite{pages-gallego_comprehensive_2023} provide valuable foundations, they \reve{have a fundamental limitation: they completely disregard RSA tools. This gap is compounded by other issues, including a lack of biological justification for design choices \omst{(e.g., training data generation)}, limited adaptability to new chemistries, and \omnd{benchmarking of tools as a whole} that obscure\omnd{s} component-level contributions}.}

\rev{Beyond \revb{the software} \omnd{benchmarks for genomic analysis}, several benchmarking \omst{framework}s have emerged to evaluate the performance \omnd{of different} genomic analysis \omnd{methods} on hardware platforms. Genomics-GPU \cite{genomics_gpu}  offers GPU-accelerated workloads for genome comparison, matching, and clustering. GenomicsBench ~\cite{genomicsbench} covers data-parallel kernels for short- and long-read workflows on CPU/GPU. While these \omst{framework}s \revb{contribute to} \omst{systematic} hardware-software co-design for genomics, they do not focus on the unique challenges of \omst{RSA}.}

\omnd{\textbf{Our goal} is to introduce an extensible benchmarking framework designed to systematically evaluate and compare text-based and RSA methods. RawBench enables: 1) a more inclusive setting towards RSA; 2) modular assessment of different reference encoding, signal encoding, and representation matching techniques; and 3) compatibility with diverse sequencing chemistries and organisms. By enabling fair, component-level evaluation, RawBench \omrd{helps and} accelerates the development of biologically informed and computationally efficient raw nanopore signal analysis methods.}

\rev{\section{RawBench Framework}\label{framework}}

\begin{figure}[tbh]
  \centering
  
  \includegraphics[width=\columnwidth]{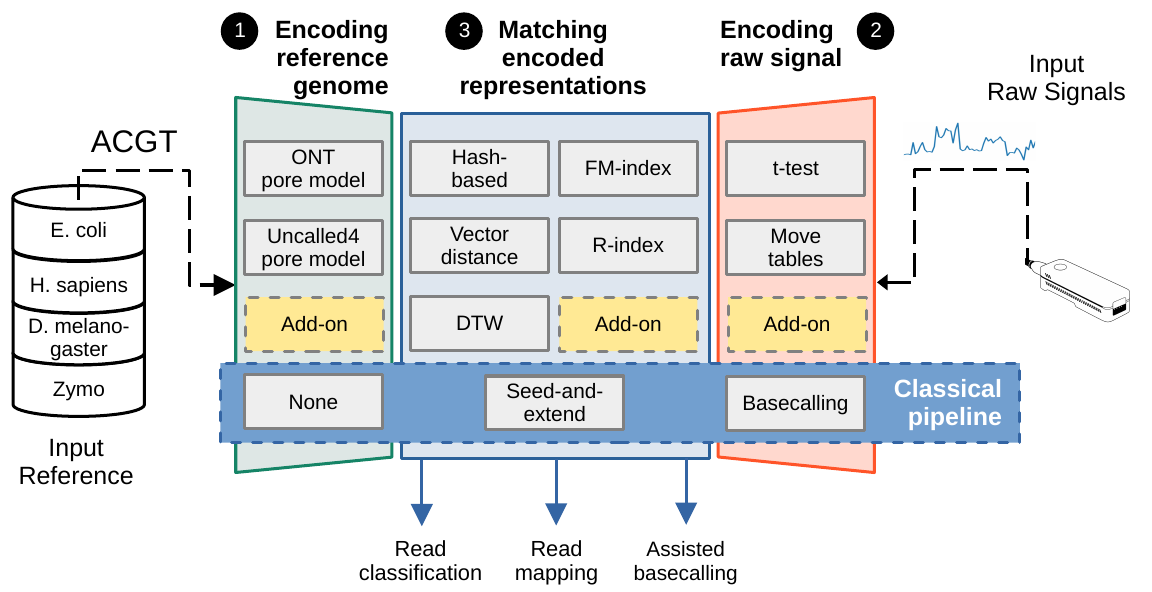}
  \caption{Overview of RawBench.}
  \label{fig:overview}
\end{figure}

\omnd{RawBench is a benchmarking framework for raw signal analysis (RSA) to evaluate RSA methods across nanopore chemistries, datasets, and computational settings, based on ground truth generated using expensive basecalled analysis. The framework is structured into three RSA stages \omrd{as shown in Figure~\ref{fig:overview}}\omrd{:} \omrd{\circlednum{1}} encoding the reference genome into expected signal patterns, \omrd{\circlednum{2}} segmenting raw signals into a comparable \omrd{encoded} representation, and \omrd{\circlednum{3}} matching these \omrd{two encoded} representations for tasks such as read mapping or classification.}

\omnd{The framework’s modular design enables as inputs (i) reference genomes, (ii) nanopore raw signals from multiple chemistries and organisms, and (iii) any RSA method that targets one or more of the three RawBench stages. To support comprehensive evaluation, RawBench includes datasets spanning bacteria, eukaryotes, and metagenomes from different nanopore chemistries, while also allowing the community to easily integrate their own datasets or downstream tasks. Beyond quality, RawBench also reports runtime and memory, providing practical insights for deploying RSA methods in real-world sequencing workflows.}

\subsection{Encoding the \omnd{R}eference \omnd{G}enome}

\reve{To enable the alignment of raw signals to their genomic origins}
\revf{directly in raw signal space,}
\reve{the reference genome must first be translated into \omst{a sequence of} expected electrical signal patterns. This is typically done using k-mer models, which map k-mers to their characteristic}
\revf{electrical}
\reve{signal distributions.}
\reve{However, these models are tightly coupled to specific nanopore chemistries and flow cell versions, limiting generalizability.}
\reve{Tools like \texttt{Uncalled4} \cite{sam_kovaka_uncalled4_2024} and \texttt{Poregen} \cite{Samarakoon2025} address this problem by learning k-mer models \textit{de novo} \omst{which} enable\omst{s faster} adaptation to new chemistries or settings.}

\reve{Despite their central role in RSA, the effectiveness of different k-mer models remains poorly understood with limited systematic benchmarking across chemistries. \omst{Factors critical for field deployment decisions} such as accuracy, memory, and runtime can vary unpredictably depending on the \omst{pore} model, underscoring the need for robust evaluation.}

\omnd{RawBench contributes by providing a unified framework to systematically benchmark k-mer models across multiple nanopore chemistries. It enables fair comparison of both official ONT and open-source learned models, measuring not only the downstream task quality but also memory and runtime trade-offs. In doing so, RawBench \omrd{enables accurate evaluation of} how reference encoding choices impact RSA quality and performance, guiding informed deployment decisions in field settings.}

\subsection{\omnd{Segmenting} the \omnd{R}aw \omnd{S}ignals}

\reve{To enable accurate alignment between raw signals and the reference genome, raw \omnd{signals} must be transformed into latent representations that meaningfully correspond to nucleotide sequences. This transformation step is central to downstream tasks such as read mapping \revf{and classification, which involves determining the origin or type of a DNA/RNA read, such as its species, gene, or functional category.} Two dominant strategies exist for this transformation.}

\reve{First, RSA requires explicit segmentation of \omst{raw} signals into discrete \omst{segments} corresponding to k-mers. \omst{Segmentation m}ethods include t-test changepoint detection~\cite{Ruxton2006} and resquiggling~\cite{gamaarachchi_gpu_2020, Simpson2017}, though these often lack awareness of sequencing context or pore-specific characteristics, limiting robustness across \crr{nanopore} chemistries.} A context-aware alternative is Campolina~\cite{bakic_campolina_2025}, a convolutional neural network (CNN) trained to predict segmentation points from raw signals. Unlike statistical methods such as t-test, which operate only on local fluctuations within a \omst{signal} chunk, Campolina leverages broader sequencing and chemistry priors to improve robustness at the cost of requiring extensive pretraining.

\reve{The second approach, basecalling, employs neural networks with CRF~\cite{10.5555/645530.655813} and CTC~\cite{10.1145/1143844.1143891} \omst{decoders} to implicitly convert signals to nucleotides. Basecallers process overlapping signal chunks that are concatenated post-processing, \omst{benefitting from a relatively long context compared to RSA approaches}. Move tables\omst{, an intermediate output} extracted from basecallers, can provide coarse segmentation points for RSA~\cite{Samarakoon2025}.} \reve{Only Bonito~\cite{oxford_nanopore_technologies_bonito_2021} and Dorado~\cite{oxford_nanopore_technologies_dorado_2024} \crr{basecallers} support \omst{the latest} R10.4.1 chemistry, but rely on proprietary training data, limiting adaptability and reproducibility.}

RawBench enables systematic evaluation of segmentation approaches by providing a wide range of benchmarks across chemis\-tries and sequencing contexts, allowing fair comparison of statistical and learning-based methods. By exposing the contrasting characteristics of different segmentation approaches, RawBench helps identify opportunities for developing future segmentation methods\omrd{, e.g., those} that combine the efficiency of statistical techniques with the robustness of learning-based models.

\subsection{Matching \omnd{E}ncoded \omnd{R}epresentations}
\label{sec:matching}

\reve{To complete the mapping process from raw signals to reference genomes, the encoded representations of both signals and reference must be matched in a way that preserves biological accuracy while maintaining computational efficiency.
Due to the high dimensionality and variability of raw signal data, designing scalable and accurate matching algorithms remains a significant challenge.}

\reve{Matching approaches generally fall into two main categories based on their computational \omst{demands} and \omst{degrees of sensitivity}: (1) heuristic methods that prioritize speed and scalability, often using hashing or indexing to perform approximate matching; and (2) precise alignment methods that emphasize \reve{quality}, typically leveraging distance metrics or dynamic programming. Each approach offers different trade-offs in terms of speed, memory, and downstream utility, especially when handling \omst{complex} genomes and high-throughput \omst{raw} signal streams. RawBench enables side-by-side evaluation of these techniques under \omst{thirty} realistic workloads.}

\reve{The first category includes hash-based and probabilistic methods. \texttt{RawHash}~\cite{firtina_rawhash_2023} and \texttt{RawHash2}~\cite{firtina_rawhash2_2024} use quantization to map similar \omst{raw} signal segments to shared hash buckets, enabling fast approximate matching for \omst{mapping against} large genomes. \texttt{UNCALLED}~\cite{kovaka_targeted_2021} employs a probabilistic FM-index to estimate the likelihood of signal-to-k-mer matches. \texttt{Sigmap}~\cite{zhang_real-time_2021} embeds both signal and reference into a shared high-dimensional space, but suffers from computational overhead and the curse of dimensionality~\cite{Bellman1966}. Compressed indexing strategies like the R-index~\cite{Gagie_1, Gagie_2} offer an alternative, supporting lightweight exact matching over repetitive regions. Tools like \texttt{Sigmoni}~\cite{Shivakumar2024} use this structure to compute pseudo-matching lengths (PMLs), where longer PMLs indicate stronger matches between compressed signal and reference segments.}

\reve{The second category focuses on fine-grained alignment methods, most notably Dynamic Time Warping (DTW) \cite{dtw_1,dtw_2}, which has been employed in earlier works to achieve high-\reve{quality} alignments between raw signals and reference sequences \cite{dunn_squigglefilter_2021, shih_efficient_2023, gamaarachchi_gpu_2020}. While DTW offers precise signal-to-reference alignment, its computational cost is prohibitive \omst{with} larger genome\omst{s}, especially in real-time scenarios. To mitigate this, hybrid approaches such as \texttt{RawAlign} \cite{lindegger_rawalign_2023} combine fast seeding techniques \cite{firtina_rawhash_2023, firtina_rawhash2_2024} with selective DTW refinement, achieving a practical trade-off between \reve{quality} and \omst{performance}.}

\reve{RawBench evaluates both types of strategies across varying genome complexities and computational demands. This extensive framework encourages rapid development and comprehensive evaluatio\crr{n} of raw signal-to-reference matching algorithms.}

\subsection{\revf{Assisting \omnd{B}asecalled \omnd{A}nalysis}} 

\crfinal{To improve efficiency in conventional basecalling pipelines, it is often desirable to avoid unnecessary basecalling of reads that are unlikely to contribute to downstream analyses. The goal of pre-basecalling \omst{raw} signal filtering\omnd{, i.e., the process of inferring raw signals that are likely to be important from signal characteristics,} is to identify such reads directly from raw signals and forward only promising reads to \omst{the basecaller}.}

By default, RSA tools process a prefix of raw signals and stop once a mapping is found to meet the constraints of real-time analysis, limiting the \omrd{amount of surrounding signal available for highly precise downstream analyses}. \crfinal{\omnd{Pre-basecalling raw signal filtering} transforms RSA into a computationally efficient preprocessing step. By \omnd{filtering out relatively insignificant} raw signals before full basecalling, \omst{it is possible to} reduce the \omst{load} on basecallers, conserve computational resources, and improve overall throughput. \omst{Filtering is performed} independently of the basecaller, making it compatible with existing pipelines without altering downstream \omst{aspects}.}

\crfinal{Tools like TargetCall~\cite{cavlak_targetcall_2024} showcase \omst{the benefits of pre-basecalling filtering} by \omst{performing an initial} lightweight \omst{raw} signal analysis, allowing \omst{for} more accurate and resource-intensive basecallers to focus on \omst{important} reads. Within RawBench, such pre-filtering approaches can be evaluated systematically across different datasets and signal complexities, enabling assessment of trade-offs between computational savings and downstream analysis quality.} \omnd{This makes RawBench a powerful tool to guide the design and deployment of future pre-basecalling strategies in practical sequencing workflows.}

\section{RawBench Datasets}

\rev{To evaluate the robustness across \omst{varying} genomic complexities, RawBench includes datasets from \textit{E. coli}, \textit{\revf{D.} melanogaster}}, \textit{\revf{H.} sapiens}, and a \textit{Zymo} metagenomic \revb{dataset} \crr{as shown in Table~\ref{tab:dataset_characteristics}. Dataset names in Table ~\ref{tab:dataset_characteristics} are provided as hypertexts referring to the source and preprocessing scripts are provided in \href{https://github.com/CMU-SAFARI/RawBench}{\faGithub}.}
 
\rev{We use nanopore sequencing data generated using the \omnd{old R9.4.1 and the} latest R10.4.1 chemistry. All datasets provide raw signals (\reb{i.e.}, FAST5 or POD5) and basecalled reads (\reb{i.e.}, FASTQ). \revf{We note that alternative formats like SLOW5 improve storage efficiency and read performance~\cite{Gamaarachchi2025}.} When possible, we \omnd{prefer} 400 bps \omnd{(i.e.,} bases per second passing through the nanopore\omnd{)} mode over the deprecated 260 bps for better quality and \omst{higher sequencing} yield~\cite{oxford_nanopore_technologies_dorado_2024}.}

\rev{Each dataset provides sufficient coverage for downstream tasks such as single nucleotide polymorphism (SNP) and structural variant (SV) calling \cite{Chaubey2020,Meynert2013,Li2011, Bizon2014}. Summary statistics are shown} in Table~\ref{tab:dataset_characteristics}.

\begin{table}[H]
\centering
\caption{\rev{Summary of RawBench datasets.}}
\label{tab:dataset_characteristics}
\small
\resizebox{\linewidth}{!}{
\begin{tabular}{@{}lcccc@{}}
\toprule
\textbf{Dataset} & \textbf{Genome Size} & \textbf{\rev{Number of}} & \textbf{\rev{Number of}} & \textbf{\rev{Depth of}}   \\
                 & \textbf{(Mbp)}       & \textbf{Reads}         & \textbf{Bases (Mbp)}       & \textbf{Coverage}       \\
\midrule
\href{https://42basepairs.com/browse/s3/human-pangenomics/submissions/5b73fa0e-658a-4248-b2b8-cd16155bc157--UCSC_GIAB_R1041_nanopore/Ecoli_R1041_Duplex_Control}{\textit{E. coli}}           & 4.6  & 180,000  & 1,131   & 225$\times$   \\
\href{https://42basepairs.com/browse/s3/ont-open-data/contrib/melanogaster_bkim_2023.01/flowcells/D.melanogaster.R1041.400bps/D_melanogaster_1/20221217_1251_MN20261_FAV70669_117da01a}{\textit{D. melanogaster}}       & 143.7  & 231,278   & 1,730  & 12.04$\times$   \\
\href{s3://ont-open-data/giab_2023.05/flowcells/hg002/20230424_1302_3H_PAO89685_2264ba8c/pod5_pass/}{\textit{H. sapiens}}            & 3,200 & 16,000  & 329   & 0.11$\times$   \\
\href{https://livejohnshopkins-my.sharepoint.com/:f:/g/personal/vshivak1_jh_edu/Ekz-z_mFGP1NqC-dlwFN58wB4feWOSvzzJmjx39N3_KSnw}{\textit{Zymo mock}}                     & 65.4 & 10,000   & 134  & 2.05$\times$  \\
\bottomrule
\end{tabular}
}
\end{table}

\revb{
RawBench\reb{ spans simple to complex genomes for evaluati\omnd{ng scalability of different RSA techniques}. \textit{E. coli} represents a compact bacterial genome for rapid testing. \textit{D. melanogaster} adds moderate complexity with repetitive and heterochromatic regions. \textit{H. sapiens} provides a highly complex genome rich in SVs, repeats (>45\%), and allelic diversity, posing the most demanding \omst{read} mapping challenge. The \textit{Zymo} mock community introduces a metagenomic case for mixed-species classification. \omst{Datasets originating from reference sequences with contrasting characteristics} ensures \omst{that} \omnd{benchmarked RSA techniques} remain robust across diverse \omst{raw nanopore sequencing data} analyses.}
}
\section{Evaluation}

\subsection{Evaluation \omnd{M}ethodology}

We implement RawBench as a modular Nextflow~\cite{DiTommaso2017} framework with C++ components, \omnd{enabling plug-and-play RSA stages from Section~\ref{framework}. In particular, stage-level techniques (e.g., t-test–based segmentation and hash-based matching) are provided as standalone C++ modules and they can be mixed and matched. For completeness, we also provide benchmarking scripts that invoke prebuilt binaries of established RSA tools; while these wrappers facilitate fair, out-of-the-box comparisons, they naturally limit full combinatorial exploration compared to our C++ modules.}

\revf{\omnd{\textbf{Methods.}} We evaluate quality, performance and coverage \omnd{of different RSA techniques}.}
To assess \omnd{read mapping and classification} quality, we create \omnd{thirty} different RSA \omnd{pipeline} combinations out of the RawBench component pool. Outputs are compared against ground truth generated by the
Dorado \cite{oxford_nanopore_technologies_dorado_2024} super-accurate (SUP) basecaller followed by \omrd{the} minimap2 \cite{li_minimap2_2018} read mapper. Metrics include true positives (TP, \crfinal{i.e.}, correctly mapped reads), false positives (FP, \crfinal{i.e.}, incorrectly mapped reads), \crfinal{false negatives (FN, \crfinal{i.e.}, reads that could not be mapped),} not aligned (NA, \crfinal{i.e.}, reads \crfinal{with no ground truth location}), precision $(\frac{TP}{TP + FP})$, recall $(\frac{TP}{TP + FN})$, and F1 score $(2 \times \frac{\text{Precision} \times \text{Recall}}{\text{Precision} + \text{Recall}})$. Results are \omnd{computed using} \texttt{UNCALLED pafstats} \omrd{\cite{kovaka_targeted_2021}}. 

For performance, we report elapsed time (wall-clock time), CPU time, and peak memory usage. GPU acceleration is \omnd{used} only for Dorado, while all RSA components (pore models, segmentation, matching) run on CPU to ensure fair comparison and reflect typical deployment scenarios.

We benchmark end-to-end \omnd{basecalled read and raw signal}  analysis while considering the impact of (1)~pore models, (2)~segmentation methods, (3)~matching methods, and (4)~RSA-assisted basecalling and 5)~basecaller context length.

\reb{Unless otherwise \omnd{stated}, evaluations fix the pore model to Uncalled4\omnd{~\cite{sam_kovaka_uncalled4_2024}}, segmentation to t-test\omnd{~\cite{Ruxton2006}}, and matching to hash-based\omnd{~\cite{Andoni2008}} (for read mapping \omrd{and RSA-assisted basecalling}) or R-index\omnd{~\cite{Gagie_1,Gagie_2}} (for read classification).} 

\omnd{\textbf{Pore Models.}} First, we evaluate how well different \emph{pore models} translate DNA into expected signal features. We isolate the effect of \revf{two} pore models\revf{, ONT\omnd{~\cite{oxford_nanopore_technologies_dorado_2024}} and Uncalled4\omnd{~\cite{sam_kovaka_uncalled4_2024}},} on downstream \reve{quality} \revb{and performance}\omrd{, corresponding to stage \rsastage{1}}.  

\omnd{\textbf{Segmentation Methods.}} Second, we assess \crcampo{three} \emph{segmentation} strategies \omnd{based} \crcampo{on their ability to partition raw signals into biologically meaningful events: 1)~t-test\omnd{~\cite{Ruxton2006}}, 2)~move tables\omnd{~\cite{oxford_nanopore_technologies_dorado_2024}}, and 3) neural network-based method Campolina~\cite{bakic_campolina_2025}}\omrd{, corresponding to stage \rsastage{2}}. 

\omnd{\textbf{Matching Methods.}} Third, we compare five different \emph{matching} algorithms to determine their effectiveness \omnd{in signal-to-reference alignment:} 1)~hash-based\omnd{~\cite{Andoni2008}}, 2)~FM-index\omnd{~\cite{Ferragina}}, 3)~vector distances\omnd{~\cite{zhang_real-time_2021}}, 4)~R-index\omnd{~\cite{Gagie_1,Gagie_2}}, and 5)~DTW\omnd{~\cite{dtw_1, dtw_2}}\omrd{, corresponding to stage \rsastage{3}}.  

\omnd{\textbf{Assisting Basecalled Analysis.}} \crr{Fourth, we evaluate RSA as a pre-filtering step\omnd{~\cite{cavlak_targetcall_2024,dunn_squigglefilter_2021,shih_efficient_2023}} for basecalling by comparing depth of coverage with and without limiting Dorado's context to successfully mapped reads from RSA.}

Fifth, we evaluate the effect of limiting Dorado\omnd{'s} context to 1, 2, 5, or 10 chunks of 4000 signal \omnd{points} each, \omnd{corresponding to approximately 1 \omrd{second} of data} \cite{firtinac_2024_rawsamble}. This reveals insights about the quality–performance trade-offs by adjusting \omnd{context length}.

\omnd{\textbf{Experimental Setup.}} \omnd{We perform a}ll experiments on a server equipped with an NVIDIA A6000 GPU \omrd{\cite{nvidia_a6000_2024}} and an Intel(R) Xeon(R) Gold 6226R CPU \omrd{\cite{intel_xeon6226R_2024}} running at 2.90 GHz. \omnd{We conduct each} evaluation with 64 threads. We show the parameters and versions of the tools we evaluate in Supplementary Tables~\ref{tab:supp_parameters}
and~\ref{tab:supp_versions} respectively.

\subsection{\revb{\reve{Quality} Evaluation}}

\omnd{Tables \ref{tab:pore_summary} and \ref{tab:pore_summary_classification} show the quality of two pore models: ONT and Uncalled4 when applied across datasets and downstream tasks. We make two key observations.}

\omnd{First, pore model choice has a noticeable impact on downstream analysis quality, with Uncalled4 outperforming ONT on read mapping tasks across all organisms \omrd{(see Table \ref{tab:pore_summary})}. Uncalled4 achieves higher F1 scores (e.g., 0.83 for \textit{E. coli}), driven largely by improvements in recall while maintaining high precision ($\geq$ 0.87). In contrast, ONT pore models consistently show lower recall, particularly in larger genomes (0.44 for \textit{H. sapiens}), limiting their ability to provide true alignments. These results \omrd{concur} with prior findings~\cite{sam_kovaka_uncalled4_2024}\omrd{,} which highlight limitations in ONT pore modeling for R10.4.1 chemistry, motivating more expressive reference-to-signal representations.}

\omnd{Second, the benefits of Uncalled4 appear to diminish in the R9.4.1 read classification task \omrd{(see Table~\ref{tab:pore_summary_classification})}. Uncalled4 and ONT pore models provide identical quality. This observation is in line with previous work~\cite{kovaka_targeted_2021}, which show\omrd{s} that ONT and Uncalled4 pore models for the old R9.4.1 chemistry exhibited almost identical characteristics. This suggests that while models like Uncalled4 can substantially improve mapping performance, they do not necessarily translate into comparable gains in higher-level tasks.}

\omnd{Together, these findings indicate that pore models remain an important determinant of read mapping quality. While advances such as Uncalled4 demonstrate clear improvements over ONT in read mapping, read classification results suggest that further research is needed to develop pore models that generalize their benefits across different downstream tasks and nanopore chemistries.}

\begin{table}[H]
\centering
\caption{\crfinal{Read mapping quality \crcampofinal{using} different pore models.}}
\label{tab:pore_summary}
\small
\begin{tabular}{l|ccc}
\hline
\textbf{Pore Model} & \textbf{F1} & \textbf{Precision} & \textbf{Recall} \\
\hline
\multicolumn{4}{c}{\textbf{\textit{E. coli}}} \\
\hline
ONT       & 0.79 & 0.88 & 0.71 \\
Uncalled4 & \textbf{0.83} & \textbf{0.91} & \textbf{0.77} \\
\hline
\multicolumn{4}{c}{\textbf{\textit{D. melanogaster}}} \\
\hline
ONT       & 0.66 & 0.93 & 0.51 \\
Uncalled4 & \textbf{0.72} & \textbf{0.94} & \textbf{0.59} \\
\hline
\multicolumn{4}{c}{\textbf{\textit{H. sapiens}}} \\
\hline
ONT       & 0.58 & 0.86 & 0.44 \\
Uncalled4 & \textbf{0.66} & \textbf{0.87} & \textbf{0.53} \\
\hline
\end{tabular}
\end{table}

\revb{\textbf{Segmentation Methods.} }
Table \ref{tab:segmentation_summary} shows the \reve{quality} of \crcampo{three} segmentation approaches: t-test, move tables\omnd{,} \crcampo{and Campolina} when applied \omrd{to} each \omnd{task} and dataset. We make \revb{two} key observations.

\revb{First, segmentation \reve{quality} has a substantial impact on downstream \omnd{task} \reve{quality}, with \crcampo{Campolina outperforming the other two approaches significantly on read mapping tasks, in particular for the larger \textit{H. sapiens} genome (F1 = 0.79).} t-test segmentation outperforms move tables across all organisms. t-test achieves high mapping F1 scores (up to 0.83 for \textit{E. coli}) and excels in classification (F1 = 0.92 on \textit{Zymo}), maintaining strong precision ($>=$ 0.85) \omnd{across all tasks}. In contrast, move tables yield\omnd{s} significantly lower \reve{quality}, likely due to \omnd{its} coarse, stride-based segmentation that introduces noise and fails to capture \omrd{many} signal transitions.}
    
\revb{Second, genomic complexity negatively affects recall, even fo\reve{r} high-performing methods like t-test \crcampo{and Campolina}. While precision remains \omrd{mostly} stable, recall drops \omnd{with larger genomes}, reflecting the challenge of accurately segmenting \omnd{signals from} complex genomes. \crcampo{This observation also holds for the neural network-based approach Campolina, albeit with some improvements in recall \omnd{for} the \textit{H. sapiens} \omnd{data}, pointing to unsolved underlying challenges in segmentation.} This trend highlights a need for segmentation approaches that retain \omnd{high} precision while improving sensitivity to capture true \omnd{segments} in \omnd{raw signals}.}

\crr{These findings show that segmentation remains a critical bottleneck: methods like t-test provide strong precision but falter on recall in complex genomes, pointing to the need for approaches that can \omnd{maintain} sensitivity without sacrificing precision.}

\begin{table}
\centering
\caption{\crfinal{Read mapping quality \omrd{using} different segmentation methods.}}
\label{tab:segmentation_summary}
\small
\begin{tabular}{l|ccc}
\hline
\textbf{Segmentation Method} & \textbf{F1} & \textbf{Precision} & \textbf{Recall} \\
\hline
\multicolumn{4}{c}{\textbf{\textit{E. coli}}} \\
\hline
t-test      & 0.83 & 0.91 & 0.77 \\
Move tables & 0.07 & 0.07 & 0.06 \\
\crcampo{Campolina} & \crcampo{\textbf{0.89}} & \crcampo{\textbf{0.94}} & \crcampo{\textbf{0.85}} \\
\hline
\multicolumn{4}{c}{\textbf{\textit{D. melanogaster}}} \\
\hline
t-test      & \textbf{0.72} & 0.94 & \textbf{0.59} \\
Move tables & 0.05 & 0.23 & 0.03 \\
\crcampo{Campolina} & \crcampo{\textbf{0.72}} & \crcampo{\textbf{0.95}}  & \crcampo{0.57}  \\
\hline
\multicolumn{4}{c}{\textbf{\textit{H. sapiens}}} \\
\hline
t-test      & 0.66 & 0.87 & 0.53 \\
Move tables & 0.01 & 0.11 & 0.01 \\
\crcampo{Campolina} & \crcampo{\textbf{0.79}} & \crcampo{\textbf{0.96}} & \crcampo{\textbf{0.67}} \\
\hline
\end{tabular}
\end{table}

\revb{\textbf{Matching Methods.}} Tables \ref{tab:matching_summary} and \ref{tab:matching_summary_classification} show the \reve{quality} of \revc{five} matching approaches: hash-based methods, FM-index, vector distances, R-index \revc{and Dynamic Time Warping (DTW)}. \revc{We note that the combination of DTW and t-test segmentation forms the basis of the f5c resquiggle method \cite{gamaarachchi_gpu_2020}.} \crr{We make \omnd{five} key observations.}

First, hash-based matching \omrd{provides} consistent\omrd{ly high} \reve{quality} across mapping tasks, \omnd{achieving} high F1 scores across organisms. This approach, used in tools like \texttt{RawHash} and \texttt{RawHash2} \cite{firtina_rawhash_2023, firtina_rawhash2_2024}, excels through its ability to quickly identify approximate matches using locality-sensitive hashing, making it computationally efficient for large-scale analyses while maintaining high \omnd{quality}.

Second, vector distance-based matching shows exceptional \reve{quality} on simpler genomes but suffers \omrd{from} dramatic \reve{quality} degradation with increasing genomic complexity. It achieves \omrd{a} high F1 score (0.83) for \textit{E. coli} mapping, comparable to hash-based approach, but \reve{quality} drops to 0.67 for \textit{D. melanogaster} and 0.26 for \textit{H. sapiens}. This decline suggests \omrd{poor scaling with} increased noise and repetitive content characteristic\reb{s} of more complex genomes.

Third, FM-index demonstrates poor \reve{quality} in \omrd{read} mapping with low F1 scores across all organisms. Consistently high precision but extremely low recall likely reflects the mismatch between exact string matching logic and the continuous nature of raw signals.

\omrd{Fourth}, mapping \reve{quality} consistently \omrd{reduces} from \textit{E. coli} to \textit{H. sapiens} across most methods, with the exception of FM-index which maintains uniformly poor \reve{quality}. Hash-based methods show the most graceful degradation, while vector distances exhibit the steepest decline. This trend underscores the scalability challenges in RSA, where increased genome complexity, repetitive content, and heterozygosity create increasingly difficult matching problems that current methods struggle to address effectively.

\omrd{Fifth}, the classification results \omrd{(see Table \ref{tab:matching_summary_classification})} reveal a different \reve{quality} landscape \omnd{in terms} of matching method effectiveness. R-index achieves perfect precision and the highest F1 score for classification, while vector distances also excel with an F1 score of 0.95. This trend indicates that while these methods struggle with fine-grained mapping, they are highly effective at organism-level discrimination tasks where approximate matches may be sufficient.

\crr{Together, these results underscore that mapping and classification favor different approaches: the former demands discriminative, fine-grained alignment, while the latter benefits from \crcampofinal{relaxed} approximate matching. This suggests that future developments in matching algorithms should consider task-specific optimizations.}

\begin{table}
\centering
\caption{\crfinal{Read mapping quality using different matching techniques.}}
\label{tab:matching_summary}
\small
\begin{tabular}{l|ccc}
\hline
\textbf{Matching Method} & \textbf{F1} & \textbf{Precision} & \textbf{Recall} \\
\hline
\multicolumn{4}{c}{\textbf{\textit{E. coli}}} \\
\hline
Hash-based       & 0.83 & 0.91 & 0.77 \\
FM-index         & 0.23 & 0.13 & 0.80 \\
Vector distances & 0.83 & 0.84 & \textbf{0.82} \\
\reb{R-index}    & 0.67 & 0.79 & 0.58 \\
\revf{DTW}       & \textbf{0.86} & \textbf{0.99} & 0.75 \\
\hline
\multicolumn{4}{c}{\textbf{\textit{D. melanogaster}}} \\
\hline
Hash-based       & 0.72 & 0.94 & 0.59 \\
FM-index         & 0.02 & 0.17 & 0.01 \\
Vector distances & \textbf{0.80} & 0.94 & \textbf{0.69} \\
\reb{R-index}    & 0.59 & \textbf{0.96} & 0.42 \\
\revf{DTW}       & 0.75 & 0.94 & 0.62 \\
\hline
\multicolumn{4}{c}{\textbf{\textit{H. sapiens}}} \\
\hline
Hash-based       & 0.66 & 0.87 & 0.53 \\
FM-index         & 0.01 & 0.05 & 0.01 \\
Vector distances & 0.26 & 0.57 & 0.16 \\
\reb{R-index}    & 0.66 & 0.85 & 0.54 \\
\revf{DTW}       & \textbf{0.75} & \textbf{0.94} & \textbf{0.62} \\
\hline
\end{tabular}
\end{table}

\begin{table}
\centering
\caption{Read classification \reve{quality} (F1, Precision, Recall) using different matching techniques.}
\label{tab:matching_summary_classification}
\small
\begin{tabular}{l | p{1cm} p{1cm} p{1cm} }
\hline
\multicolumn{4}{c}{\textbf{\textit{Zymo}}} \\
\hline
\textbf{Matching Method} & \textbf{F1}  & \textbf{Precision} & \textbf{Recall}  \\
\hline
Hash-based           & 0.95 &  0.92 & 0.97  \\
FM-index             & 0.62 &  0.45 & \textbf{0.99}  \\
Vector distances       & 0.96 & 0.97 & 0.95 \\
R-index             & 0.96 &  \textbf{1.0} & 0.93 \\
\revf{DTW}                 & \textbf{0.98} & 0.99 & 0.97 \\
\hline
\end{tabular}
\end{table}

\textbf{Assisting Basecalled Analysis.}
Table~\ref{tab:rsa_assisted_rmaq} shows coverage \revf{results} comparing \omrd{RSA-}assisted versus \omrd{RSA-}unassisted basecalling across two organisms. We make \omnd{two} key observations.

\omrd{First, RSA assistance lowers average depth of coverage while keeping breadth nearly unchanged. In E. coli, depth drops from 184.04× to 164.39×, yet breadth stays at ~82.4\%. In D. melanogaster, depth decreases from 15.24× to 11.61× with breadth remaining ~99.9\%. This trend shows that RSA filtering reduces redundant coverage and basecalling while maintaining completeness.}

Second, genom\omnd{e} complexity determines the effectiveness of \omnd{RSA assistance}. \textit{D. melanogaster} maintains nearly identical breadth coverage despite 39\% fewer reads (99.82\% vs 99.91\%) \omrd{with RSA assistance} \omnd{which suggests that lower-\reve{quality} reads are filtered \omrd{out}, yielding sensitive but fewer mappings.} \textit{E. coli} shows more modest improvements with 17\% fewer reads and minimal breadth change (82.42\% vs 82.49\%), suggesting that RSA assistance is more beneficial with complex genomes.

The\omrd{se} results demonstrate that RSA \omrd{assistance in basecalling} provides \omnd{substantial} benefits for \omnd{complex} genomes by substantially reducing the basecalling load. Although we exclude \textit{\revd{\revf{H.}} sapiens} due to performance overheads \omrd{incurred by the RSA pre-filtering}, results suggest that tailored RSA tools could be developed to enable \omrd{RSA assistance} benefits for organisms with larger genomes, potentially across the full spectrum of genomic complexity.

\begin{table}
\centering
\caption{Basecalled read mapping \reve{quality} analysis.}
\label{tab:rsa_assisted_rmaq}
\small
\setlength{\tabcolsep}{3pt}
\begin{tabular}{@{}lcccc@{}}
\hline
\textbf{Dataset} & \textbf{\crr{RSA}} & \textbf{Average Depth} & \textbf{Breadth of} & \textbf{Aligned}  \\
 & \textbf{\omrd{Pre-filter}} & \textbf{of Cov. ($\times$)} & \textbf{Coverage (\%)} & \textbf{Reads (\#)}  \\
\hline
\textbf{\textit{E. coli}} & \revf{\CheckmarkBold}           & 164.39 & 82.42 & 182,871 \\
& \revf{\XSolidBrush}              & \omrd{\textbf{184.04}} & \omrd{\textbf{82.49}} & \omrd{\textbf{221,651}} \\
\hline
\textbf{\textit{D. melanogaster}} & \revf{\CheckmarkBold}           & 11.61 & 99.82 & 152,601 \\
& \revf{\XSolidBrush}              & \omrd{\textbf{15.24}} & \omrd{\textbf{99.91}} & \omrd{\textbf{251,868}} \\
\hline
\end{tabular}
\end{table}

\omrd{Next, w}e examine the effect of limited context length on basecallers to evaluate their benefits in real-time analysis where a portion of raw signals is basecalled rather than the entire raw signal. Table \ref{tab:basecallers_limited_chunks} shows the \reve{quality} of Dorado SUP model with varying context length\omrd{s} in terms of number of chunks of \omnd{raw signal points} across three organisms. We make \omnd{two} key observations.

First, genom\omnd{e} complexity determines sensitivity to context \omrd{length} limitations. \textit{D. melanogaster} demonstrates the strongest response to increased context \omrd{length}, with TP rates improving from 78.30\% to 94.10\%  and FP rates decreasing from 20.40\% to 5.31\% as context \omrd{length} increases. In contrast, \textit{E. coli} shows modest improvements 
while \textit{H. sapiens} exhibits the least amount of gains. These results indicate that additional context enables \omrd{better} basecalling decisions, but its benefits diminish for \omrd{\textit{H. sapiens}} where extra context may not resolve ambiguities in homopolymeric regions.

\omnd{Second}, Read Until\reb{~\cite{loose_real-time_2016}} applications can achieve substantial \reve{quality} gains with relatively modest increases in context length, particularly from 1 to 5 chunks. \omnd{Largest benefits} occur in the \omrd{first few context length increases}, with diminishing returns beyond \omrd{the first} 5 chunks. This finding indicates that Read Until strategies could \omrd{strike} \omnd{a} balance between performance and \reve{quality} by using adaptive context lengths, especially given that \textit{D. melanogaster} achieves 89.85\% TP with 5 chunks compared to 94.10\% with 10 chunks.

The findings suggest that adaptive context length strategies, tailored to the expected genomic complexity of the target organism, could optimize Read Until sequencing \reve{quality} while managing computational overhead. \omrd{T}he substantial improvements observed with even modest increases in context length (from 1 to 2-5 chunks) indicate that small increases in sequencing time can yield significant gains in basecalling \reve{quality}\omrd{, for some organisms}.

\begin{table}[H]
\centering
\caption{Effect of limited number of chunks on Dorado (SUP) basecaller.}
\label{tab:basecallers_limited_chunks}
\small
\begin{tabular}{lccccc}
\hline
\textbf{Dataset} & \textbf{Chunks (\#)} & \textbf{TP} & \textbf{FP} & \textbf{NA}  \\
\hline
\hline
\textbf{\textit{E. coli}}&1             & 84.50 & 2.90 & 12.60 \\
&2             & 84.99 & 2.54 & 12.46 \\
&5             & 85.77 & 2.15 & 12.07 \\
&10           & \omrd{\textbf{88.47}} & \omrd{\textbf{1.24}} & \omrd{\textbf{10.29}} \\
\hline
\hline
\textbf{\textit{D. melanogaster}}&1              & 78.30 & 20.40 & 1.30 \\
&2              & 83.46 & 15.40 & 1.14 \\
&5              & 89.85 & 9.23 & 0.92 \\
&10           & \omrd{\textbf{94.10}} & \omrd{\textbf{5.31}} & \omrd{\textbf{0.59}} \\
\hline
\hline
\textbf{\textit{H. sapiens}}&1             & 80.67 & 10.80 & 8.52 \\
&2             & 80.56 & 11.12 & \omrd{\textbf{8.32}} \\
&5             & 81.89 & 9.50 & 8.61 \\
&10           & \omrd{\textbf{84.07}} & \omrd{\textbf{7.32}} & 8.61 \\
\hline
\end{tabular}
\end{table}

\subsection{\revb{Performance Evaluation}}

\revb{\textbf{Pore Models.}}  
Table~\ref{tab:mapping_pore_performance} summarizes the \omrd{read mapping }performance of different pore models. We make \omnd{two} key observations.  

First, Uncalled4 provides substantial runtime benefits for smaller and intermediate genomes. For \textit{E. coli} and \textit{D. melanogaster}, Uncalled4 reduces elapsed time by \omrd{2}0–\omrd{5}0\% relative to ONT, while also reducing CPU time and memory footprint. These improvements suggest that the more compact Uncalled4 pore representation accelerates signal processing without sacrificing quality, consistent with the quality results in Table~\ref{tab:pore_summary}.  

Second, performance benefits \omrd{of Uncalled4} diminish for larger genomes. For \textit{H. sapiens}, Uncalled4 \omnd{enables} faster \omnd{execution} with a higher peak memory demand. This indicates a shift in bottlenecks, where reduced computational overhead is offset by increased memory usage, again likely due to the denser signal lookup tables created using Uncalled4 when applied to complex genomes ~\cite{sam_kovaka_uncalled4_2024}.  

These results highlight the importance of aligning pore model choice with both dataset scale and available system resources.  

\begin{table}[H]
\centering
\caption{\reb{Read mapping performance \omrd{using} different pore models.}}
\label{tab:mapping_pore_performance}
\small
\begin{tabular}{l|cccccc}
\hline
\textbf{Pore Model} & \textbf{Elapsed time}  & \textbf{CPU time} & \textbf{Peak}  \\
\                           & \textbf{(hh:mm:ss)}   & \textbf{(sec)}    & \textbf{Mem. (GB)} \\\toprule
\multicolumn{4}{c}{\textbf{\textit{E. coli}}} \\
\hline

ONT           & 0:08:22 &  6\crcampofinal{,}481 & 4.46\\
Uncalled4             & \omrd{\textbf{0:05:51}} & \omrd{\textbf{5,730}} & \omrd{\textbf{4.36}} \\
\hline
\multicolumn{4}{c}{\textbf{\textit{D. melanogaster}}} \\
\hline

ONT           & 2:41:07 &  596,614 & 10.25\\
Uncalled4             & \omrd{\textbf{2:07:02}} & \omrd{\textbf{462,608}} & \omrd{\textbf{9.6}} \\
\hline

\multicolumn{4}{c}{\textbf{\textit{H. sapiens}}} \\
\hline

ONT           & 1:46:47 &  189,846 & \omrd{\textbf{80.02}}\\
Uncalled4             & \omrd{\textbf{0:53:03}} & \omrd{\textbf{186,301}} & 91.96 \\
\hline
\end{tabular}
\end{table}

\revb{\textbf{Segmentation Methods.}}  
\crr{Tables~\ref{tab:mapping_segmentation_performance} and ~\ref{tab:classification_segmentation_performance} show the performance of different segmentation methods. We note one key pattern.}

\crr{Move tables, precomputed by the Dorado (SUP) basecaller, are supplied externally to the RSA pipeline. To ensure fairness, we report results including the computation time for move tables. In this setting, move tables come with additional GPU runtime, while offering no improvement over the simpler t-test segmentation. This suggests that move tables are not yet optimized for RSA. However, targeted refinement of move tables for \omnd{raw signal segmentation}, similar to their utility in other contexts~\cite{Samarakoon2025,bakic_campolina_2025,ont_remora}, could enable them to outperform current methods in future iterations.}

\revb{\textbf{Matching Methods.}}  
Table~\ref{tab:mapping_matching_performance} summarizes the performance of different matching methods. We \omnd{make} \omrd{three} \omnd{key observations}.  

First, R-index \omnd{offers} \omrd{a} balance \omrd{between} speed and memory \omnd{footprint}. Across all organisms, R-index delivers the fastest execution while keeping memory demands low. This trend favors the use of R-index for real-time and large-scale analyses.  

Second, vector distance methods \omrd{are} computationally demanding. While achieving strong \omnd{quality} (Table~\ref{tab:matching_summary}), they require orders of magnitude more memory and runtime—up to six hours and 265 GB for the \textit{H. sapiens} dataset. This resource intensity makes \omrd{FM-index a practical alternative } for complex genomes despite \omrd{the headroom for} \omnd{quality}.

\omrd{Third}, hash-based and DTW matching occupy intermediate positions with distinct trade-offs. Hash-based matching is more memory-efficient, while DTW, by contrast, requires more memory but substantially less than vector distances, with runtimes that scale gracefully with genome size.  

\omrd{The results highlight that no single method dominates: R-index is fastest on \textit{E. coli}, DTW on \textit{D. melanogaster}, and FM-index leads on \textit{H. sapiens} and in peak memory demand across organisms. Vector distances methods are consistently the most memory-intensive. The changing trends indicate that method choice should be made based on genome complexity and resource limitations.}

\begin{table}
\centering
\caption{Read mapping performance \omrd{using} different matching methods.}
\label{tab:mapping_matching_performance}
\small
\begin{tabular}{l|cccccc}
\hline
\textbf{Matching Method} & \textbf{Elapsed time}  & \textbf{CPU time} & \textbf{Peak}  \\
\                           & \textbf{(hh:mm:ss)}   & \textbf{(sec)}    & \textbf{Mem. (GB)} \\\toprule
\multicolumn{4}{c}{\textbf{\textit{E. coli}}} \\
\hline

Hash-based           & 0:05:51 &  5,730 & 4.36  \\
FM-index             & 6:57:45 &  1,603,653 & \omrd{\textbf{1.09}}  \\
Vector distances       & 0:20:10 & 54,310 & 54.32 \\
\revf{R-index}                 & \omrd{\textbf{0:04:25}} & \omrd{\textbf{4,224}} & 1.4 \\
\revf{DTW}                 & 0:09:23 & 6,128 & 4.43 \\\hline
\multicolumn{4}{c}{\textbf{\textit{D. melanogaster}}} \\
\hline
Hash-based           & 2:07:02 &  462,608 & 9.6  \\
FM-index             & 3:56:13 &  892,824 & \omrd{\textbf{1.49}}  \\
Vector distances       & 3:22:15 & 823,117 & 255.97 \\
\revf{R-index}                 & 1:22:30 & 310,695 & 3.1 \\
\revf{DTW}                 & \omrd{\textbf{0:24:02}} & \omrd{\textbf{88,044}} & 10.46 \\
\hline
\multicolumn{4}{c}{\textbf{\textit{H. sapiens}}} \\
\hline
Hash-based           & 0:53:03 &  186,301 & 91.96  \\
FM-index             & \omrd{\textbf{0:08:44}} &  \omrd{\textbf{32,808}} & \omrd{\textbf{7.52}}  \\
Vector distances       & 5:59:29 & 1,238,190 & 265.16 \\
\revf{R-index}                 & 0:35:02 & 131,095 & 29.43 \\
\revf{DTW}                 & 0:46:35 & 158,289 & 116.2 \\\hline
\end{tabular}
\end{table}
\section{Resources}
\label{resources}
The RawBench framework, including datasets, evaluation scripts, and documentation, is available on \revf{\href{https://github.com/CMU-SAFARI/RawBench}{\url{https://github.com/CMU-SAFARI/RawBench}}}. \crr{The framework is} publicly available to facilitate its adoption and extension by the research community.

\section{Conclusion}

RawBench provides a comprehensive benchmarking \omst{framework} for \omrd{raw signal analysis (RSA)} that addresses critical gaps in current frameworks. By decomposing analysis pipelines into three modular components (i.e., reference genome encoding, signal encoding, and representation matching) and evaluating them across organisms of varying genomic complexity, we demonstrate that \crfinal{statistical segmentation methods outperform the intermediate outputs of basecallers, though carefully designed context-aware ML models can surpass these statistical approaches. We further show that advanced pore models like Uncalled4 offer consistent \reve{quality} improvements and hash-based matching provides the most robust quality across genome complexities}. The framework's modular design enables systematic evaluation of emerging methods while maintaining biological relevance \omrd{via} up-to-date datasets. RawBench establishes a foundation for systematic progress in \omrd{RSA}, enabling researchers to design more effective pipelines and unlock the full potential of raw signal data for diverse genomics applications.

\section*{\crcampofinal{Acknowledgments}}

\noindent \crcampofinal{We thank the anonymous reviewers of ACM BCB 2025 for their valuable feedback. We thank the SAFARI \omrd{Research} \omrd{G}roup members for thoughtful feedback and the \omst{stimulating} intellectual \omrd{\&} scientific environment they cultivate. We acknowledge the generous support of our industrial partners, including Google, Huawei, Intel, and Microsoft. This work \omrd{is} partially supported by the ETH Future Computing Laboratory (EFCL) \omst{and} the European Union’s Horizon research and innovation programme [101047160 - BioPIM].}

\bibliographystyle{unsrtnat}
\balance
\bibliography{references}

\appendix
\onecolumn
\setcounter{secnumdepth}{3}
\clearpage
\begin{center}
\textbf{\LARGE Supplementary Material for\\ RawBench: A comprehensive benchmarking framework for raw nanopore signal analysis}
\end{center}
\setcounter{section}{0}
\setcounter{equation}{0}
\setcounter{figure}{0}
\setcounter{table}{0}
\setcounter{page}{1}
\renewcommand{\thetable}{S\arabic{table}}
\renewcommand{\thefigure}{S\arabic{figure}}

\section{Extended Quality Benchmarks}
\label{suppsec:accuracy_benchmarks}

\begin{table}[H]
\centering
\caption{Read classification \reve{quality} (F1, Precision, Recall) \omrd{using} different pore models.}
\label{tab:pore_summary_classification}
\small
\begin{tabular}{l|cccccc}
\hline
\multicolumn{4}{c}{\textbf{\textit{Zymo}}} \\
\hline
\textbf{Pore Model} &  \textbf{F1} & \textbf{Precision} & \textbf{Recall} \\
\hline
ONT           & \omrd{\textbf{0.96}} & \omrd{\textbf{1.0}} & \omrd{\textbf{0.93}}  \\
Uncalled4             & \omrd{\textbf{0.96}} & \omrd{\textbf{1.0}} & \omrd{\textbf{0.93}} \\
\hline
\end{tabular}
\end{table}

\begin{table}[H]
\centering
\caption{Read classification \reve{quality} (F1, Precision, Recall) for different segmentation methods.}
\label{tab:segmentation_summary_classification}
\small
\begin{tabular}{l | p{1cm} p{1cm} p{1cm} p{1cm} p{1cm} p{1cm}}
\hline
\multicolumn{4}{c}{\textbf{\textit{Zymo}}} \\
\hline
\textbf{Segmentation Method} & \textbf{F1}  & \textbf{Precision} & \textbf{Recall}  \\
\hline
t-test & \textbf{0.96} & \textbf{1.0} & \textbf{0.93} \\
Move tables      & 0.20 & 0.24 & 0.17\\
\hline
\end{tabular}
\end{table}

\section{Extended Performance Benchmarks}
\label{suppsec:performance_benchmarks}
\omrd{\noindent\textbf{Note.} \emph{NA} in Tables~\ref{tab:mapping_segmentation_performance} and ~\ref{tab:classification_segmentation_performance} indicates the metric was not applicable for that run.}

\begin{table}[H]
\centering
\caption{Read classification performance \omrd{using} different pore models.}
\label{tab:classification_pore_performance}
\small
\begin{tabular}{l|cccccc}
\hline
\textbf{Pore Model} & \textbf{Elapsed time}  & \textbf{CPU time} & \textbf{Peak}  \\
\                           & \textbf{(hh:mm:ss)}   & \textbf{(sec)}    & \textbf{Mem. (GB)} \\\toprule

\multicolumn{4}{c}{\textbf{\textit{Zymo}}} \\
\hline

ONT           & \omrd{\textbf{0:03:42}} &  \omrd{\textbf{792}} & \omrd{\textbf{2.53}}\\
Uncalled4             & 0:04:31 & 1,085 & \omrd{\textbf{2.53}} \\
\hline
\end{tabular}
\end{table}

\begin{table}[H]
\centering
\caption{\reb{Read mapping performance \omrd{using} different segmentation methods.}}
\label{tab:mapping_segmentation_performance}
\small
\begin{tabular}{l|cccccc}
\hline
\textbf{Segmentation Method} & \textbf{Elapsed time}  & \textbf{CPU time} & \textbf{Peak}  \\
\                           & \textbf{(hh:mm:ss)}   & \textbf{(sec)}    & \textbf{Mem. (GB)} \\\toprule

\multicolumn{4}{c}{\textbf{\textit{E. coli}}} \\
\hline

t-test          & \omrd{\textbf{0:05:51}} & 5,730 & \omrd{\textbf{4.36}}\\
Move tables             & 0:38:52 & NA & 14.64 \\

\hline

\multicolumn{4}{c}{\textbf{\textit{D. melanogaster}}} \\
\hline

t-test           & 2:07:02 & 462,608 & \omrd{\textbf{9.6}}\\
Move tables             & \omrd{\textbf{1:39:47}} & NA & 18.25 \\

\hline

\multicolumn{4}{c}{\textbf{\textit{H. sapiens}}} \\
\hline

t-test           & \omrd{\textbf{0:53:03}} & 186,301 & \omrd{\textbf{91.96}}\\
Move tables             & 0:53:26 & NA & 93.18 \\

\hline
\end{tabular}
\end{table}

\begin{table}[H]
\centering
\caption{Read classification performance \omrd{using} different segmentation methods.}
\label{tab:classification_segmentation_performance}
\small
\begin{tabular}{l|cccccc}
\hline
\textbf{Segmentation Method} & \textbf{Elapsed time}  & \textbf{CPU time} & \textbf{Peak}  \\
\                           & \textbf{(hh:mm:ss)}   & \textbf{(sec)}    & \textbf{Mem. (GB)} \\\toprule

\multicolumn{4}{c}{\textbf{\textit{Zymo}}} \\
\hline

t-test           & \omrd{\textbf{0:04:31}} &  \omrd{\textbf{1,085}} & \omrd{\textbf{2.53}}\\
Move tables             & 0:12:57 & NA & 3.95 \\
\hline
\end{tabular}
\end{table}

\begin{table}[H]
\centering
\caption{Read classification performance \omrd{using} different matching methods.}
\label{tab:classification_matching_performance}
\small
\begin{tabular}{l|cccccc}
\hline
\textbf{Matching Method} & \textbf{Elapsed time}  & \textbf{CPU time} & \textbf{Peak}  \\
\                           & \textbf{(hh:mm:ss)}   & \textbf{(sec)}    & \textbf{Mem. (GB)} \\\toprule

\multicolumn{4}{c}{\textbf{\textit{Zymo}}} \\
\hline

Hash-based           & \omrd{\textbf{0:00:40}} &  312 & 4.29  \\
FM-index             & 0:02:30 &  1,855 & \omrd{\textbf{0.88}}  \\
Vector distances       & 0:01:30 & 2,347 & 10.2 \\
\revf{R-index}                 & 0:04:31 &  1,085 & 2.53 \\
\revf{DTW}                 & \omrd{\textbf{0:00:40}} & \omrd{\textbf{289}} & 4.66 \\\hline
\end{tabular}
\end{table}

\section{Configuration} \label{suppsec:configuration}
\subsection{Parameters} 

In Supplementary Table~\ref{tab:supp_parameters}, we show the details of the parameters used for each tool and dataset including preset values, when required. For minimap2~\cite{li_minimap2_2018}, we use the same parameter setting for all datasets. For \omrd{the} Dorado super-accurate (SUP) basecaller, we use the model trained for the corresponding data sampling frequency (i.e. 4 kHz or 5 kHz). Thread count was specified as 64 \omrd{for all CPU workloads, i.e., all processes other than basecalling which used a GPU}. 
\begin{table}[tbh]
\centering
\caption{Parameters we use in our evaluation for each tool and dataset in mapping.}
\label{tab:supp_parameters}
\begin{tabular}{@{}lccc@{}}\toprule
\textbf{Tool} & \textbf{\emph{E. coli}} & \textbf{\emph{D. melanogaster}} & \textbf{\emph{H. sapiens}} \\\midrule
Minimap2          & \multicolumn{3}{c}{-x map-ont}\\\midrule
Dorado GPU (SUP) & \multicolumn{3}{c}{basecaller dna\_r10.4.1\_e8.2\_400bps\_sup@\textbf{v4.1.0}/\textbf{4.1.0}/\textbf{5.0.0} --emit-moves}\\\midrule
RawHash2 & \multicolumn{3}{c}{-r10 -x sensitive} \\\midrule
Sigmoni & \multicolumn{3}{c}{-b 6 --shred 100000 --complexity --thresh 1.6666666666333334}\\\bottomrule
\end{tabular}
\end{table}
\label{suppsubsec:parameters}
\subsection{Versions}\label{suppsubsec:versions}

Supplementary Table~\ref{tab:supp_versions} shows the version and the link to these corresponding versions of each tool we use in our experiments. \omrd{Scripts to reproduce each of the experiments can be found on \href{https://github.com/CMU-SAFARI/RawBench}{\url{https://github.com/CMU-SAFARI/RawBench}}.}

\begin{table}[tbh]
\centering
\caption{Versions of each tool and library.}
\label{tab:supp_versions}
\begin{tabular}{@{}lll@{}}\toprule
\textbf{Tool} & \textbf{Version} & \textbf{Link to the Source Code} \\\midrule
RawHash2 & 2.1 & \url{https://github.com/CMU-SAFARI/RawHash/releases/tag/v2.1}\\\midrule
Minimap2 & 2.28-r1209 & \url{https://github.com/lh3/minimap2/releases/tag/v2.28}\\\midrule
Dorado & 0.9.6 & \url{https://github.com/nanoporetech/dorado/releases/tag/v0.9.6}\\\midrule
Sigmoni &  {} & \url{https://github.com/vikshiv/sigmoni}\\\midrule
UNCALLED & 2.3 & \url{https://github.com/skovaka/UNCALLED/releases/tag/v2.3}\\\midrule
Uncalled4 & 4.1.0 & \url{https://github.com/skovaka/uncalled4/releases/tag/4.1.0}\\\midrule
Sigmap & 0.1 & \url{https://github.com/haowenz/sigmap/releases/tag/v0.1}\\\midrule
Mosdepth & 0.3.10 & \url{https://github.com/brentp/mosdepth/releases/tag/v0.3.10}\\\midrule
samtools & 1.22 & \url{https://github.com/samtools/samtools/releases/tag/1.22}\\\midrule
bedtools & 2.31.1 & \url{https://github.com/arq5x/bedtools2/releases/tag/v2.31.1}\\\bottomrule
\end{tabular}

\end{table}

\let\noopsort\undefined
\let\printfirst\undefined
\let\singleletter\undefined
\let\switchargs\undefined

\end{document}

%% file: commands.tex
\definecolor{denim}{rgb}{0.08, 0.38, 0.74}
\definecolor{darkolivegreen}{rgb}{0.33, 0.42, 0.18}
\definecolor{dgreen}{rgb}{0.00, 0.75, 0.00}
\definecolor{darkpink}{rgb}{0.88, 0.28, 0.54}
\definecolor{forestgreen}{rgb}{0.0, 0.27, 0.13}
\definecolor{amber}{rgb}{1.0, 0.49, 0.0}
\definecolor{lightyellow}{rgb}{0.980, 0.956, 0.623}
\definecolor{lightblue}{rgb}{0.980, 0.956, 0.623}
\definecolor{darkamber}{rgb}{0.5, 0.19, 0.0}
\definecolor{dkgreen}{rgb}{0,0.6,0}
\definecolor{gray}{rgb}{0.5,0.5,0.5}
\definecolor{mauve}{rgb}{0.58,0,0.82}
\definecolor{lightmauve}{rgb}{0.68,0.4,0.92}
\definecolor{chocolate}{rgb}{0.48, 0.25, 0.0}
\definecolor{dollarbill}{rgb}{0.52,0.73,0.4}
\definecolor{dkdkgreen}{rgb}{0,0.45,0}
\definecolor{gfored}{rgb}{0.580, 0.050, 0.211}
\definecolor{darkwarmgray}{rgb}{0.15, 0.050, 0.05}
\definecolor{ups-truck}{rgb}{0.53, 0.28, 0.21}
\definecolor{darktangerine}{rgb}{1.0, 0.66, 0.07}

\definecolor{bestresult}{HTML}{b3e2cd} 
\definecolor{worseresult}{HTML}{f4c7c3}

\newcommand\rev[1]{{\color{black}{#1}}}
\newcommand\revb[1]{{\color{black}{#1}}}
\newcommand\revc[1]{{\color{black}{#1}}}
\newcommand\revd[1]{{\color{black}{#1}}}
\newcommand\reve[1]{{\color{black}{#1}}}
\newcommand\revf[1]{{\color{black}{#1}}}

\newcommand\omst[1]{{\color{black}{#1}}}
\newcommand\omnd[1]{{\color{black}{#1}}}
\newcommand\omrd[1]{{\color{black}{#1}}}

\newcommand\reb[1]{{\color{black}{#1}}}
\newcommand\crr[1]{{\color{black}{#1}}}
\newcommand\crfinal[1]{{\color{black}{#1}}}
\newcommand\crcampo[1]{{\color{black}{#1}}}
\newcommand\crcampofinal[1]{{\color{black}{#1}}}

\makeatletter
\g@addto@macro{\normalsize}{
  \setlength{\abovedisplayskip}{2pt plus 1pt minus 1pt}
  \setlength{\belowdisplayskip}{2pt plus 1pt minus 1pt}
  \setlength{\intextsep}{2pt plus 1pt minus 1pt}
  \setlength{\textfloatsep}{3pt plus 1pt minus 1pt}
  \setlength{\dbltextfloatsep}{3pt plus 1pt minus 1pt}
  \setlength{\skip\footins}{4pt plus 1pt minus 1pt}
}
\def\BibTeX{{\rm B\kern-.05em{\sc i\kern-.025em b}\kern-.08em
    T\kern-.1667em\lower.7ex\hbox{E}\kern-.125emX}}

\newcommand{\squishlist}{
 \begin{list}{$\circ$}
  { \setlength{\itemsep}{0pt}
     \setlength{\parsep}{0pt}
     \setlength{\topsep}{0pt}
     \setlength{\partopsep}{0pt}
     \setlength{\leftmargin}{1em}
     \setlength{\labelwidth}{1em}
     \setlength{\labelsep}{0.5em} } }

\newcommand{\squishsublist}{
\begin{list}{$\rightarrow$}
 { \setlength{\itemsep}{0pt}
    \setlength{\parsep}{0pt}
    \setlength{\topsep}{-10em}
    \setlength{\partopsep}{-3pt}
    \setlength{\leftmargin}{1em}
    \setlength{\labelwidth}{1em}
    \setlength{\labelsep}{0.5em} } }

\newcommand{\squishend}{
  \end{list}  }